\documentclass[aps,prb,twocolumn,showpacs,amssymb,amsmath]{revtex4}
\usepackage[dvips]{graphicx,color}
\usepackage{color}
\begin{document}

\title {Anomalous suppression of the shot noise in a
nanoelectromechanical system} 

\author{Federica Haupt$^{1}$, Fabio Cavaliere$^{1}$, Rosario
Fazio$^{2,3}$, and Maura Sassetti$^{1}$
\vspace{1mm}} \affiliation{ $^{1}$ Dipartimento di Fisica,
Universit\`a di Genova and LAMIA-INFM-CNR, Via Dodecaneso 33, 16146
Genova, Italy\\ $^{2}$ International School for Advanced Studies
(SISSA), Via Beirut 2-4, 34014 Trieste, Italy\\ $^{3}$ NEST-INFM-CNR
and Scuola Normale Superiore, Piazza dei Cavalieri 7, 56126 Pisa,
Italy\\
\vspace{3mm}} \date{July 3, 2006}
\begin{abstract} 
In this paper we report a relaxation--induced suppression of the noise
for a single level quantum dot coupled to an oscillator with
incoherent dynamics in the sequential tunneling regime. It is shown
that relaxation induces qualitative changes in the transport
properties of the dot, depending on the strength of the
electron-phonon coupling and on the applied voltage. In particular,
critical thresholds in voltage and relaxation are found such that a
suppression below $1/2$ of the Fano factor is possible. Additionally,
the current is either enhanced or suppressed by increasing relaxation,
depending on bias being greater or smaller than the above
threshold. These results exist for any strength of the
electron--phonon coupling and are confirmed by a four states toy
model.
\end{abstract}
\maketitle
\noindent
\section{Introduction}
In the last years, nanoelectromechanical systems (NEMS) have been a
hot research topic both from the theoretical and the experimental
point of view. ~\cite{craighead,cleland} Combining electronic and
mechanical degrees of freedom, NEMS have potentially important
applications as fast and ultra--sensitive
detectors,~\cite{cleland1,knobel,ekinci,schwab} as well as being
interesting dynamical systems in their own right. In these devices,
current can be used both to create and to detect vibrational
excitations.  Clear evidence of phonon excitations induced by single
electron tunneling has been reported in a number of different systems,
including semiconducting phonon cavities,~\cite{weig}
molecules~\cite{park,pasupathy} and suspended carbon
nanotubes.~\cite{leroy,sapmaz} At finite bias electrons tend to drive
phonons out of equilibrium; signatures of non equilibrium phonon
distribution were observed in a suspended carbon
nanotube.~\cite{leroy}

On the theoretical side, NEMS are often described with simple
phenomenological models involving a single electron device coupled to
an harmonic oscillator.
~\cite{boese,mccarthy,nowack,gorelik,novotny,pistolesi,flensberg, armour,
kochPRB,mitra,koch,belzig,blanter,usmani,jauho,martin,kochFCS} Even
within these simple models, many peculiar features such as negative
differential conductance,~\cite{boese,mccarthy,nowack} shuttling
instability~\cite{gorelik,novotny} and strong mechanical
feedback~\cite{blanter,usmani}have been predicted in the case of an
underdamped oscillator. It is then a physically relevant question
whether the vibrational energy is reduced by relaxation processes
induced by coupling to an external environment~\cite{flensberg} or
rather because tunneling itself.~\cite{armour,kochPRB,mitra} Up to
now, theoretical works have focused mostly on the case of negligible
relaxation, taking the opposite case of strong relaxation as a
reference term. Significant differences between these two cases have
been found both for weak and strong electron--phonon (e--ph) coupling.~\cite{mitra,koch}

Many recent theoretical works have focused on the study of current
noise on NEMS.~\cite{armour, mitra,koch,belzig,blanter,usmani,jauho}
In particular, the Fano factor $F$, which is the ratio between the
zero frequency component of the noise and the average current, has
proven to be very sensitive to the e--ph interaction
and to the details of the phonon distribution.~\cite{mitra, koch} A
giant enhancement of Fano factor ($F \sim 10^2-10^3$) has been
predicted for strong coupling and negligible phonon
relaxation.~\cite{koch} In the opposite limit of strong relaxation,
i.e.  when the phonons are thermally distributed, a NEMS behaves
essentially as a single electron transistor (SET).~\cite{flensberg} 
Shot noise in SET has been extensively studied~\cite{buttiker,korotkov,loss}
and $F$ was always found to be larger or equal to $1/2$.
However, Fano factors  below this limit in the single electron tunneling regime have
been predicted in more complicated systems. For istance, coupling to internal degrees
of  freedom~\cite{egues} can induce $F$ slightly below 1/2  ($F \sim 0.45$). 
A strong suppression of the noise has been predicted for the quantum
shuttle.~\cite{jauho,pistolesi}
In this case, very low values of the Fano factors stem from an highly
ordered charge transfer mechanism given by strong correlations between charge and
mechanical motion. 

In this work we discuss how intermediate phonon relaxation influence the
transport properties of a SET coupled to a mechanical oscillator.
We focus on the sequential tunneling
regime and we adopt a rate equation to describe the dynamics of the
system. This approach is justified when the characteristic
frequency of the oscillator is much bigger than the tunneling
rate,~\cite{boese,mccarthy,mitra,martin} which is the typical
experimental situation. We find that finite relaxation rate affects
the dynamics in a highly non trivial way. Both current
and noise can be either enhanced or suppressed by relaxation,
depending on the e--ph coupling and on the considered voltage
range. In particular, for voltages higher than a certain critical
value, the Fano factor can be even {\em suppressed below} $1/2$. This
suppression is observed in a completely incoherent regime as a
consequence of the interplay between vibration assisted tunneling and
direct relaxation of different vibrational states.\\ 

The paper is organized as follows. The model Hamiltonian is defined in
Sec.~\ref{II}, while in Sec.~\ref{III} we introduce the rate equation
and the formal expressions for the current and the noise. In
Sec.~\ref{IV} numerical results for the current and Fano factor are
presented: in particular, the suppression of the Fano factor is
discussed in detail for a wide range of parameter.  Finally, analytic
expressions for the current and the Fano factor are derived within a
toy model employing few phononic states.

\section{Model}  \label{II}
In several experimental realizations, either using lithographically
defined quantum dots,~\cite{knobel,weig}
molecules~\cite{park,pasupathy} or nanotubes,~\cite{leroy,sapmaz}
electron transport is dominated by single electron
tunneling~\cite{ingold}.  In this regime, the system is essentially a
SET coupled to an harmonic oscillator. Describing the SET as a single
electronic level, the Hamiltonian of the system is
$H_s=H_n+H_b+H_{n,b}$ where~\cite{koch,mitra} ($\hbar=1$)
\begin{eqnarray}
\label{H_S}
&H_{n\phantom{,n}}&=\varepsilon \, n, \\
&H_{b\phantom{,n}}&=\omega_{0}\,(b^{\dagger}b+1/2),\\
&H_{n,b}&=\lambda\,\omega_{0}\,(b^{\dagger}+b)\,n.
\end{eqnarray}
The operator $n=d^{\dag}d$ represents the occupation number of the
single level, whose energy $\varepsilon=\varepsilon(V_g)$ can be tuned
with the aid of an external gate voltage $V_g$.  Vibrational
excitations are created by $b^{\dagger}$ and their ground state is
defined as the zero--phonon state when $n=0$. The frequency of the
oscillator $\omega_0$ can range from the hundreds of MHz of a
nanometrical cantilever~\cite{roukes} to a dozen of THz in the case of
molecular devices or suspended nanotubes.~\cite{pasupathy,leroy} The
dimensionless parameter $\lambda$ in the coupling term $H_{n,b}$
represents the strength of the e--ph interaction. For example,
$\lambda \sim 1$ was reported for the $\rm{C}_{60}$
devices~\cite{park} and for suspended carbon nanotubes,~\cite{sapmaz}
while values of $\lambda$ between 0.4 and 3 have been found in
different ${\rm C}_{140}$ samples.~\cite{pasupathy}

The SET is coupled to external leads by a tunneling Hamiltonian
\begin{equation}
\label{eq:tunham}
H_t=\sum_{k,\alpha=1,2}t_{\alpha}(c^{\dag}_{k,\alpha}d+d^{\dag} c_{k,
\alpha}),
\end{equation}
where the operators $c_{k,\alpha}^{\dagger}$ create electrons
with momentum $k$ in lead $\alpha=1,2$. The leads are described as non
interacting Fermi liquids with
\begin{equation}
\label{eq:leads}
H_{leads}=\sum_{k,\alpha=1,2}\varepsilon_{k,\alpha}c^{\dagger}_{k,\alpha
} c_{k,\alpha}
\end{equation}
and their chemical potential can be shifted by a bias voltage $V$. For
simplicity, in the following we will assume symmetric voltage drops
and symmetric barriers $t_1=t_2\equiv t_0$.

Finally, the oscillator is  coupled a dissipative environment that we
describe as a set of harmonic oscillators~\cite{caldeiralegget}
\begin{eqnarray}
&H_{env}&=\sum_{j}\omega_{j}(a^{\dagger}_{j}a_{j}+1/2), \\
&H_{b,env}&=\sum_{j}\chi_{j}\omega_{j}(a^{\dagger}_{j}+a_{j})(b^{\dagger}+b).
\end{eqnarray}
Here $a_j^{\dag}$ are the creation operators of the bosonic bath
modes. The environmental coupling is usefully characterized by its
spectral function
\begin{equation} \label{Jomega}
\mathcal{J}(\omega)=2 \pi\sum_j \omega_j^2 \chi_j^2 \delta(\omega-\omega_j).
\end{equation}


\section{Rate Equation} \label{III}
The eigenstates of $H_s$ can be written as
$|n,l\rangle$, where $n$ denotes the occupation of the single level
and $l$ the phonon number.  The coupling to the leads and to the
environment induces an energy broadening of these eigenstates. If this
broadening is the smallest energy scale of the problem, a perturbative
treatment for $H_{t}$ and $H_{b,env}$ is appropriate and a master
equation for the reduced density matrix of the system can be derived
in the sequential tunneling regime.~\cite{breuer} At lowest order, the
reduced density matrix is diagonal in $n$ but may
still be off--diagonal in $l$ because of the e--ph coupling $H_{n,b}$.

In the following we will consider the case where $\omega_0$ is much larger
than the bare tunneling rate $\Gamma^{(0)} =2 \pi \nu t_{0}^2$ (with
$\nu $ the density of states of the leads). 
In this ``diabatic" regime, which is the typical experimental
situation,~\cite{park,pasupathy,leroy} the elements of the density matrix that
are non--diagonal in phonon number become
negligible.~\cite{mitra,mccarthy,martin,boese} Then, the master equation
reduces to a rate equation for the occupation probabilities $P_{nl}$ of
the state $|n,l\rangle$
\begin{eqnarray} 
\label{ME}
\frac{d}{d t} P_{nl} &=&\sum_{n'\neq n} \sum_{l',\alpha}[P_{n'l'} {\Gamma_{\! \!\scriptstyle \alpha}}_{\,l' \to l}^{\,n' \to n}-P_{nl} {\Gamma_{\! \!\scriptstyle \alpha}}_{\,l \to l'}^{\,n \to n'}]\nonumber\\
&+&\sum_{l'}[P_{nl'} \Gamma_{l' \to l}^{rel}-P_{nl} \Gamma_{l \to l'}^{rel}].
\end{eqnarray}
The coefficients ${\Gamma_{\! \!\scriptstyle \alpha}}_{\,l \to
l'}^{\,n \to n'}$ represent the tunneling rates through the $\alpha$-th barrier
while $\Gamma_{l \to l'}^{rel}$ are  the relaxation rates.
 
In order to evaluate such rates, it is convenient to eliminate the
coupling term $H_{n,b}$ from $H_{s}$ by means of a canonical
transformation. Due to the coupling term $H_{b,env}$, this
transformation must include both the operators of the oscillator and
those of the environment~\cite{flensberg}
\begin{equation*}
\bar{O}=e^{A n}Oe^{-A n},\quad A=\kappa (b^{\dag}-b)-2 \kappa \sum_j \chi_j
(a^{\dag}_j-a_j),
\end{equation*}
where
\begin{equation}
\kappa=\frac{\lambda}{1-4 \sum_j \chi_j^2 \omega_j/\omega_0}.
\end{equation}
The total Hamiltonian is transformed into
\begin{equation*}
\bar{H}=\bar{H}_{n}+H_{b}+H_{leads}+H_{env}+\bar{H}_{t}+H_{b,env}
\end{equation*}
where 
\begin{equation} \label{Htbar}
\bar{H}_t=\sum_{k, \alpha=1,2} t_{0}(c^{\dag}_{k, \alpha}
e ^{-A}d+d^{\dag} e^{A} c_{k, \alpha}).
\end{equation}
and $\bar{H}_{n}=\bar{\varepsilon}\, n$ with
$\bar{\varepsilon}=\varepsilon- \lambda \kappa \omega_0$.  As
the energy of the SET is renormalized by a factor proportional to
$\lambda^2$, this represents the relevant parameter for the e--ph
interaction.

The transition rates can now be easily calculated using Fermi golden
rule, giving rise to tunneling rates proportional to $t_{0}^2$, and
relaxation rates, which depend on $\chi_j^2$.

The relaxation rates represent transitions between vibrational
excitations without change of the electronic state ($\beta^{-1}=k_B
T$)
\begin{equation} \label{relax}
\Gamma_{l\to (l-1)}^{rel}=  e^{\beta \omega_0 } \Gamma_{(l-1) \to
l}^{rel}=l\frac{\mathcal{J}(\omega_{0})}{1-e^{-\beta \omega_0 }},
\end{equation}
where $\mathcal{J}(\omega_{0})$ is the spectral density of the phonon
bath  Eq.~(\ref{Jomega}), evaluated at the frequency of the
oscillator. Treating $H_{b,env}$ at second order allows only
transitions between neighboring states (i.e. $|l'-l| =1$). Transitions
with $|l'-l| \ge 1$ can be included considering different relaxation
mechanisms.~\cite{boese,koch}

The charge transfer rates are induced by $\bar{H}_{t}$. Assuming the
electrons in the leads are at equilibrium with their chemical
potential, one obtains the following expressions
\begin{eqnarray} 
{\Gamma_{\! \!\scriptstyle \alpha}}_{\,l \to l'}^{0 \to 1}&=&
\Gamma^{(0)}
X_{l'l} f_{\alpha}(\omega_0(l'-l)), \label{tunnel-in} \\
{\Gamma_{\! \!\scriptstyle \alpha}}_{\,l \to l'}^{1 \to 0}&=&
\Gamma^{(0)}
X_{l'l} [1-f_{\alpha}(\omega_0(l-l'))], \label{tunnel-out}
\end{eqnarray}
where $f_{\alpha}(x)\equiv f(x+\bar{\varepsilon}-
\delta\mu_{\alpha})$, $f(x)$ is the Fermi function and
$\delta\mu_{\alpha}=(-1)^{\alpha+1}eV/2$ is the shift of the chemical
potential of the leads induced by the bias voltage.  The coefficients
$X_{ll'}$ are given by
\begin{eqnarray}
\label{FC}
X_{ll'}&=&|\langle n,l|e^{-\lambda
(b^{\dag}-b)}|n,l'\rangle|^2\nonumber\\
&=&e^{-\lambda^2}{\lambda}^{2|l-l'|}
\frac{l_{\scriptscriptstyle{<}}!}{l_{\scriptscriptstyle{>}}!} \big|
L_{l_{\scriptscriptstyle{<}}}^{|l-l'|}(\lambda^2) \big|^2,
\end{eqnarray}
where $l_{\scriptscriptstyle{<}}={\rm min}\{l,l'\}$,
$l_{\scriptscriptstyle{>}}={\rm max}\{l,l'\}$ and $L_{l}^{n}(x)$ is a
generalized Laguerre polynomial. These terms are called Franck--Condon
factors and are well known from molecular
spectroscopy.~\cite{herzberg} The effect of the e--ph interaction on
transport is two--fold: on one hand it suppresses the effective
tunneling rate (because of the factor $e^{-\lambda^2}$), on the other
it induces a non--trivial dependence on the phononic indices $l,l'$.
Up to moderate e--ph coupling ($\lambda^2 \le 1$), transitions which
conserve or change slightly $l$ have the largest
amplitude and those between states with low vibrational number  are
dominant.  Vice versa, the latter are exponentially suppressed for
$\lambda^2 \gg 1$, while transitions which change $l$ considerably
become favored.

Within the rate equation approach, the current and noise can be
evaluated by means of standard techniques.~\cite{ingold,korotkov} It
is convenient to adopt a matrix formalism and write the rate equation
as
\begin{equation} \label{master}
|\dot{P} \rangle=\mathcal{M}|P\rangle,
\end{equation}
where the vector $|P \rangle \equiv \{ P_{nl} \}$ represents the time dependent
occupation probability distribution. Calling $|{P}^{(st)}\rangle$ the
stationary solution of Eq.~(\ref{master}), the steady current through
the $\alpha$-th barrier is
\begin{equation} \label{I}
\langle I_{\alpha}\rangle=e\, \langle {\bf 1}|\mathcal{I}_{\alpha}| P^{(st)}\rangle, \quad
\alpha={1,2}
\end{equation}
where  $\langle{\bf 1}| \equiv (1,1,\dots,1)$, $e$ is the the charge of the electron 
and $\mathcal{I}_{\alpha}$ is a matrix representing all the possible
transitions through the considered barrier
$(\mathcal{I}_{\alpha})_{l,\, l'}^{n,n'}\! \!\equiv\!\!
(-1)^{\alpha+1}(n-n')\, {\Gamma_{\! \!  \scriptscriptstyle
    \alpha}}_{\,l' \to l}^{n' \to n}$.
Following Ref.~\cite{korotkov}, the zero frequency component of the
noise is given by
\begin{eqnarray}
S_{\alpha,\beta}(0) &=& -2 e^2\langle {\bf 1}|\big( \delta\mathcal{I}_{\alpha}
\mathcal{M}^{-1} \delta\mathcal{I}_{\beta}
+ \delta\mathcal{I}_{\beta}
\mathcal{M}^{-1} \delta\mathcal{I}_{\alpha} \big)
|P^{(st)}\rangle \nonumber  \\ 
&\phantom{=}& + 2 e^2 \,\delta_{\alpha,\beta}\, \langle
{\bf 1}|\,|\mathcal{I}_{\alpha}|\,|
P^{(st)}\rangle\,,
\end{eqnarray}
where $\delta\mathcal{I}_{\alpha}=\mathcal{I}_{\alpha}-\langle
I_{\alpha} \rangle/e$.
Because of charge conservation, the steady current and the zero
frequency current correlators are independent of barrier index:
$\langle I_{\alpha}\rangle=I$, $S_{\alpha, \beta} (0)= S$.
In the following we will mostly refer to the Fano factor $ F=S/2eI $.

\section{Results} \label{IV}
\subsection{Full solution of the rate equation}
The dynamics of the system is characterized by two competing time
scales: the average time spent by an electron in the dot $\tau_{el}$
and the phonon relaxation time $\tau_{ph}$.
If $\tau_{el}\gg\tau_{ph}$, the vibrational excitations tend to relax
between each tunneling event to the thermal Bose distribution
$P_l^{(eq)}=e^{-\beta l \omega_0}(1-e^{-\beta \omega_0})$.  In this
limit, charge and vibrational degrees of freedom decouple $P_{nl}=P_n
P_l^{(eq)}$ and the dynamics of the system reduces to an effective
two--state sequential tunneling process.~\cite{flensberg} The analytic
expressions for current and noise are well known~\cite{buttiker} and,
for $k_BT \ll eV$, are respectively given by
\begin{equation}
I^{(eq)}=e\,\frac{\tilde{\Gamma}_1\tilde{\Gamma}_2}{\tilde{\Gamma}_1+
\tilde{\Gamma}_2},\quad \quad  
F^{(eq)}=\frac{\tilde{\Gamma}_1^2+\tilde{\Gamma}_2^2}{(\tilde{\Gamma}_1+
\tilde{\Gamma}_2)^2}.
\end{equation} 
Here $\tilde{\Gamma}_{1}= \Gamma^{(0)}\sum_l a_{l} f_{1}(l\omega_0)$ and
$\tilde{\Gamma}_{2}= \Gamma^{(0)}\sum_l a_{l}[1- f_{2}(-l\omega_0)]$ are
the renormalized rates for tunneling in and out of the dot and $a_l$
are Poissonian weight factors~\cite{weights} $a_l = \theta(l)
e^{-\lambda^2}\lambda^{2l}/l!$. In this case the smallest possible
value of the Fano factor is $F^{(eq)}= 1/2$.

Vice versa, if $\tau_{el}\ll\tau_{ph}$ the tunneling electrons drive
the vibrations out of equilibrium and peculiar features such as
negative differential conductance (NDC)~\cite{boese,nowack,martin} and
super--Poissonian shot--noise~\cite{koch} have been predicted.

In our model, a rough estimate of $\tau_{el}^{-1}$ is given by
 $\tau_{el}^{-1}= \Gamma^{(0)} e^{-\lambda^2} $, i.e. by the effective
 transparency of the barrier set by the e--ph coupling, while
 $\tau_{ph}^{-1}$ is determined by environment spectral density
 $\tau_{ph}^{-1}=\mathcal{J}(\omega_0)$.  It is useful to define a
 dimensionless parameter for the relaxation strength
\begin{equation}
\label{defw}
w=\mathcal{J}(\omega_{0})/\Gamma^{(0)}\,.
\end{equation}
In terms of $w$, the condition for equilibrated phonons $\tau_{el} \gg
\tau_{ph}$ reads $w \gg \exp(-\lambda^2)$. 
It is then evident that the e--ph coupling defines
a characteristic scale for relaxation: the stronger is the coupling,
the more sensitive is the system to phonon relaxation.

\begin{figure}[htpb]
 \setlength{\unitlength}{1cm}
  \includegraphics[clip,width=8.0cm,keepaspectratio]{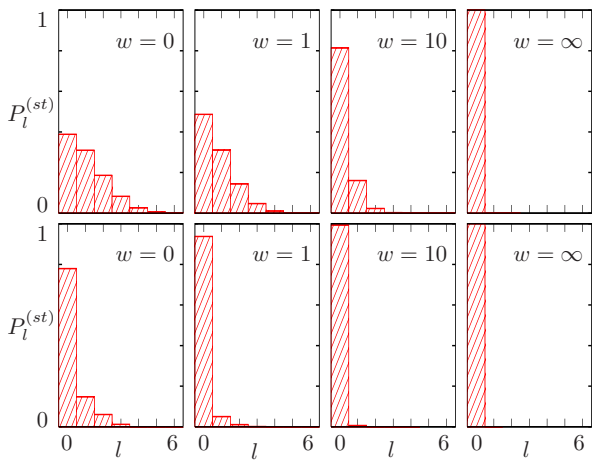}
  \vskip-0.1cm
 \caption[]{Stationary phonon probability distribution $P^{(st)}_{l}$ for different
values of $w$. Upper panels:
     $\lambda^2=0.4$, lower: $\lambda^2=7$. The rightmost panels
     ($w=\infty$) represent the thermal Bose distribution
     $P^{(eq)}_l=e^{-\beta l \omega_0}(1-e^{-\beta \omega_0})$.
     Other parameters: $V=3\ \omega_0$, $\bar{\varepsilon}=0$ and $k_B
     T=0.02\ \omega_0$.}
 \label{Fig.relax}
\end{figure}

This is reflected by the stationary phonon distribution
$P^{(st)}_{l}=P_{0\,l}^{(st)}+P_{1\,l}^{(st)}$. For increasing
relaxation strength $w$, $P^{(st)}_{l}$ tends monotonically to
$P^{(eq)}_{l}$ but with $\lambda^2$--depending speed (see
Fig.\ref{Fig.relax}). For strong e--ph coupling, the phonon
distribution is narrow already in the non--relaxed case~\cite{mitra}
$w=0$ and it reaches equilibrium for values of $w$ which are sensibly
smaller than for weak $\lambda^2$.

\begin{figure}[htpb]
 \setlength{\unitlength}{1cm}
  \includegraphics[clip,width=8.0cm,keepaspectratio]{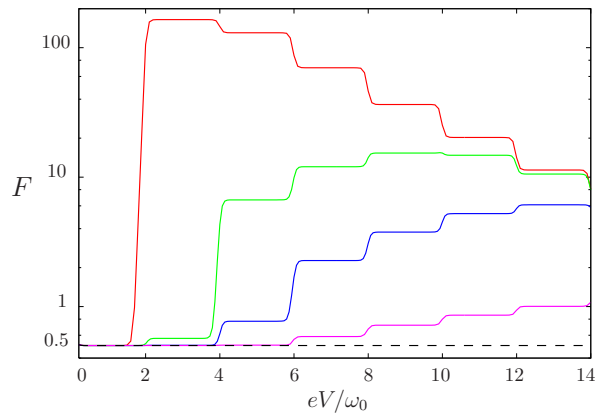}
  \vskip-0.1cm
 \caption[]{Fano factor as a function of
   voltage for $\lambda^2=16$ and for different values of the relaxation
   strength $w$; red $w=0$, green $w=0.1$, blue $w=1$, magenta $w=10$.
   Dashed line, $F=1/2$. Other parameters: $\bar{\varepsilon}=0$
   and $k_B T=0.02\, \omega_0$.}
\label{superF}
\end{figure}

Since $P^{(st)}_{l}$ converges monotonically to $ P_l^{(eq)}$ for
growing $w$, one expects most of the features of the non equilibrated
case to be washed out by increasing relaxation.  This is particularly
evident in the case of the giant Fano factor observed at low voltages
for strong interaction ($\lambda^2 \gg 1$), which is strongly
suppressed even by weak relaxation (see Fig.~\ref{superF}). This
behavior can be easily understood observing that $F \gg 1$ depends
dramatically on the non equilibrium distribution of the
vibrational excitations induced by tunneling.~\cite{koch,kochFCS}  In
fact, for large $\lambda^2$ transitions between low lying phonon
states are exponentially suppressed (see
Eq.~(\ref{FC})). Therefore, the main contribution to the current
comes from high excited vibrational states (states with large $l$) but
at low voltages the occupation probability of those states is strongly
suppressed.~\cite{mitra,nowack} These conditions leads to avalanches
of tunneling processes which, in turn, are responsible for the huge
values of $F$.~\cite{koch,kochFCS} Direct phonon relaxation inhibits
this mechanism reducing even further the occupation of states with
large $l$ and, consequently, both the current and the Fano factor are
strongly suppressed. For very strong relaxation ($w \to \infty$),
$F\to1/2$ as one would expect for equilibrated phonon on resonance
($\bar{\varepsilon}=0$).

Similarly, relaxation has a destructive effect on NDC (not shown) as
this is also a consequence of the peculiarity of the nonequilibrium
phonon distribution induced by tunneling itself.~\cite{nowack,martin}

One could be tempted to conclude that considering explicitly the
effects of relaxation simply results in an ``interpolating" behavior
between the opposite limits of no relaxation and thermally distributed
phonons. However, we find that \emph{finite} relaxation rate can
induce unexpected features.

\begin{figure}[htpb]
 \setlength{\unitlength}{1cm}
  \includegraphics[clip,width=8.0cm,keepaspectratio]{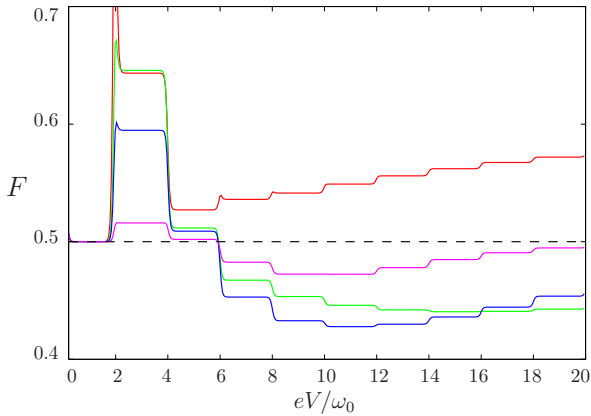}
  \vskip-0.1cm
 \caption[]{Fano factor as a function of
   voltage at $\lambda^2=3$ and for different values of the relaxation
   strength $w$; red $w=0$, green $w=5$, blue $w=12$, magenta $w=100$.
   Dashed line, $F= 1/2$. Other parameters: $\bar{\varepsilon}=0$
   and $k_B T=0.02\, \omega_0$.}
\label{Feta3}
\end{figure}

Let's first consider the case of moderate coupling $\lambda^2=3$. 
In Fig.~\ref{Feta3} we plot the Fano factor as a function of
voltage for different values of $w$ and for $\bar{\varepsilon}=0$.
It appears that $F$ has a non systematic dependence on $w$: it can
be either enhanced or suppressed by relaxation depending on the considered
voltage range. For $eV<6\omega_0$ it is always $F \ge 1/2$. In particular,
for $eV<2\omega_0$ it is  $F = 1/2$ as the tunneling electrons cannot
excite vibrations and the system behaves as an ordinary single level. 
More interestingly, for $eV>6\omega_0$, relaxation can suppress
$F$ even \emph{below} $1/2$. 
 
It is worthwhile to stress that we are dealing with a single
electronic level in the sequential tunneling regime, coupled to a
harmonic oscillator with a completely incoherent dynamics. Therefore,
this unexpected suppression of the Fano below 1/2 can only be ascribed
to the interplay between vibration--assisted tunneling and direct
relaxation of the phononic excitations. Indeed, relaxation induces a
tendency to ordered transfer of electrons thought the SET via
emission--absorption of phonons.

\begin{figure}[htpb]
 \setlength{\unitlength}{1cm}
  \includegraphics[clip,width=8.5cm,keepaspectratio]{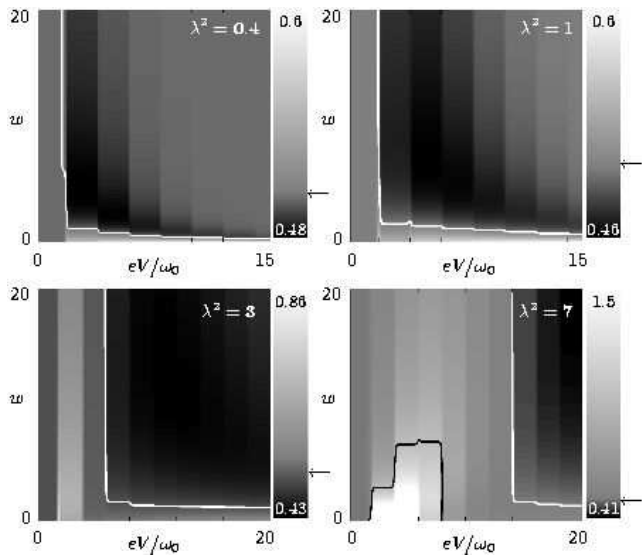}
  \vskip-0.1cm
 \caption[]{Density plot of the Fano factor as a function of bias $V$ and
   relaxation $w$ for different values of $\lambda$.  In all the
   panels: dark gray $F<1/2$, medium gray $F=1/2$ (indicated by the
   arrow in the color map ) and light gray $F>1/2$.  The white line,
   corresponding to $F=1/2$, represents $w_t(V)$. The black line in
   the 4th panel delimits the region where noise is superpoissonian,
   $F > 1$. Other parameters: $\bar{\epsilon}=0$ and $k_B T=0.02\,
   \omega_0$.}
 \label{density}
\end{figure}
From a numerical analysis, it emerges that this peculiar behavior can
be found for {\em any} value of the e--ph coupling. In particular, we
observed that it exists a voltage threshold set by the e--ph
interaction
\begin{equation}\label{vth}
eV_{t}(\lambda) = 2\omega_0\ \mathrm{int}[\lambda^2] 
\end{equation}
such that, for $V>V_{t}(\lambda)$, relaxation larger than a certain
threshold value $w_{t}(V,\lambda)$ suppresses the Fano factor below
$1/2$. This is shown in Fig.~\ref{density}, which represents a
grayscale plot of the Fano factor in the $(V,w)$-plane, for different
values of $\lambda^2$. The white contour line corresponds to $F=1/2$
and separates two different regions in the $(V,w)$-plane: the one to
the right of the contour, where $F<1/2$ and the other one where
$F>1/2$. In other words, the white line denotes $w_t$ as a function of
$V$ at given $\lambda^2$. The threshold voltage $V_t(\lambda)$
corresponds to the position of the vertical asymptote of
$w_{t}(V,\lambda)$.  For $\lambda^2<2$ the critical voltage coincides
with the onset of vibration assisted tunneling
$eV_t(\lambda)=2\omega_0$; vice versa for strong e--ph
coupling ($\lambda^2 \gg 1$), $V_t(\lambda)$ becomes very large and  this
is why $F$ is always higher than $1/2$ in Fig.~\ref{superF}.

\begin{figure}[htpb]
 \setlength{\unitlength}{1cm}
  \includegraphics[clip,width=8.0cm,keepaspectratio]{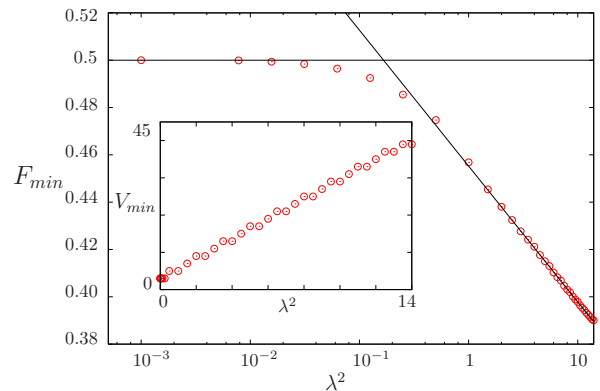}
  \vskip-0.1cm
 \caption[]{Main panel: $F_{min}$ as a function of $\lambda^2$. Each point
corresponds to different 
   values of $w$ and $V$. Inset: voltage $V_{min}$ where the minimum
   is found as a function of $\lambda^{2}$.}
 \label{Fmin}
\end{figure}

The minimal value assumed by the Fano factor $F_{min}$ depends itself
on the e--ph coupling (see Fig.~\ref{Fmin}). For weak coupling,
$F_{min}$ differs only slightly from 1/2. For stronger coupling
($\lambda^2>1$) it decreases logarithmically and it only reaches the
value $F_{min}\sim 0.4$ for considerably strong interactions. Note
that each point in Fig.~\ref{Fmin} corresponds to different values of
voltage and relaxation strength, as the position of $F_{min}$ in the
$(V,w)$-plane depends on $\lambda^2$.  The inset shows the voltage
$V_{min}$ where the minimum is found.

\begin{figure}[htpb]
 \setlength{\unitlength}{1cm}
  \includegraphics[clip,width=8.0cm,keepaspectratio]{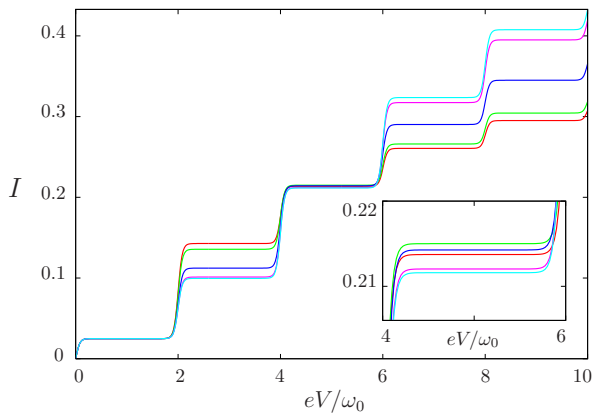}
  \vskip-0.1cm
 \caption[]{Current as a function of voltage for $\lambda^2=3$ and for 
different values of the relaxation strength $w$; red $w=0$, green $w=1$, blue
$w=10$, magenta $w=100$, cyan $w=\infty$. Other parameters 
$\bar{\varepsilon}=0$ and $k_B T=0.02 \omega_0$. Inset: zoom of the plateau
around $eV=5 \omega_0$. Current in units $e \Gamma^{(0)}$.}
\label{current}
\end{figure}

Finally, let's observe that for $\lambda^2>2$, the threshold voltage
$V_t(\lambda)$ corresponds to the onset of the transition $l:0 \to
\mathrm{int}[\lambda^2]$. In this case $V_t(\lambda)$ is a
characteristic voltage also for the current which can be either
suppressed or enhanced by relaxation depending on $V$ being smaller or
larger than $V_t(\lambda)$ (see Fig.~\ref{current}).  Relaxation
contributes to populate the low lying phonon states and then, at low
voltages, it inhibits the current as the transitions between those
states have exponentially suppressed rates. However, for
$V>V_{t}(\lambda)$ the transition $l:0 \to \mathrm{int}[\lambda^2]$ is
allowed and, as it correspond the greatest Franck--Condon
factor,~\cite{nowack} it gives a substantial contribution to the
current. In this case relaxation has the opposite effect and it
sustains the current ``feeding'' the population of the vibrational
ground state. For $V \sim V_t(\lambda)$, these two mechanisms coexist
and, consequently, the current depends only  weakly on relaxation
(see inset in Fig.~\ref{current}).

This observation fits nicely  what is reported in
literature.~\cite{koch,mitra} In fact, for $\lambda^2<2$ the critical
voltage is smaller than the energy required to have phonon--assisted tunneling
and then the current 
is enhanced by phonon relaxation at any voltage,
consistently to what was observed in Ref.~\cite{mitra} Vice versa for
very strong e--ph coupling the enhancement of the current due to relaxation can be hardly seen~\cite{koch} as  $V_t(\lambda)$ shifts to very large
voltages.
\subsection{Toy model}
To get some insight in these results, we focus on the low voltage
region $eV<4 \omega_0$ and we consider a toy model with only four
accessible state, i.e. $n={0,1}$ and $l={0,1}$. In this case
analytical expressions for current and noise can be derived. For sake
of simplicity, we report only the solutions on resonance
($\bar{\varepsilon}=0$) and at zero temperature
\begin{equation}\label{I4}
I=e\Gamma^{(0)}\Big[\frac{X_{00}}{2}+\theta (eV-2 \omega_0) \frac{X_{01}(w+2 X_{01}-\Delta)}{2
(w+2 X_{01})}\Big],
\end{equation}
and 
\begin{equation}\label{F4}
F=\frac{1}{2}-
\frac{\theta (eV-2 \omega_0) X_{01} \Delta [w^2+w(2 X_{01}-\Delta)-
X_{01}\Delta]}{(w+2 X_{01})^2K},
\end{equation}
where $\Delta=X_{00}-X_{11}$ and $K=[w (X_{01}+X_{11}+\Delta)+X_{01}(2
X_{01}+2 X_{11}+\Delta)]$.  

From Eq.~(\ref{I4}) it is easy to show that the current is an
increasing function of $w$ only for $\Delta>0$ (that is, for
$\lambda^2<2$, see Eq.~(\ref{FC})). Vice versa, for $\Delta<0$
($\lambda^2>2$) the current decreases for increasing relaxation, in
agreement with what was previously observed. Moreover Eq.~(\ref{F4})
tells that $\Delta>0$ ($\lambda^2<2$) is the necessary condition to
have $F<1/2$ in the region $2\omega_0<eV<4 \omega_0$. In fact only in
this case, it exist a threshold value for relaxation
\begin{equation}
2w_{t}(\lambda)=\Delta-2X_{01}+\sqrt{\Delta^2+4X_{01}^{2}},
\end{equation}
such that for $w>w_t$ the Fano factor is smaller than 1/2. For
stronger e--ph coupling $\lambda^2>2$ ($\Delta<0$), it is always
$F>1/2$. This confirms the numerical estimate
$eV_{t}(\lambda)=2\omega_0$ as the threshold voltage for any
$\lambda^2<2$.

\begin{figure}[htpb]
 \setlength{\unitlength}{1cm}
  \includegraphics[clip,width=8.0cm,keepaspectratio]{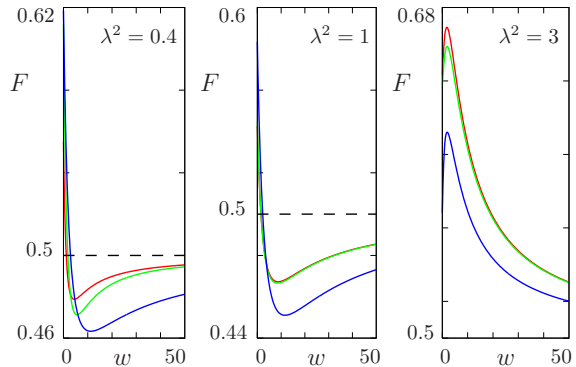}
  \vskip-0.1cm
 \caption[]{Fano factor as a function of relaxation strength $w$
   for $eV=3\omega_0$ and for different values of the e--ph interaction.
   Red curve: exact numerical result; blue curve: result for the four
   states model - Eq.~(\ref{F4}); green curve: result for a six states
   model with $n=\{0,1 \}$ and $l=\{ 0,1,2 \}$. Other parameters
   $\bar{\epsilon}=0$ and $k_B T=0.02 \omega_0$.}
 \label{scanrel}
\end{figure}

Despite the coarseness of the model, Eq.~(\ref{F4}) accords
qualitatively with the exact numerical solution for $eV< 4\omega_0$
(see Fig.~\ref{scanrel}). The agreement is reasonably good even in the
case of weak e--ph coupling, where the phonon distribution is mostly
broadened~\cite{mitra} and one expects the four state approximation to
be more inaccurate. A better agreement can be obtained considering a
six state model with $n=0,1$ and $l=0,1,2$ but, in this case, the
analytic solutions become quite cumbersome and we don't report them
here for simplicity. The agreement of the four states model with
numerical result suggests that $F<1/2$ rather depends on the interplay
between relaxation and vibration--assisted tunneling, than on the
possibility to access an high number of vibrational states.

\section{Conclusions}
\label{conclusions}
In conclusion, we have investigated the effects of  direct phonon
relaxation on the shot noise of a SET coupled to a mechanical
oscillator. For increasing relaxation strength, the occupation 
probability distribution of the states of the
system  evolves {\em monotonically}  towards thermal equilibrium. In
contrast, we found a {\em non--monotonous} behavior of the Fano
factor, which can be {\em suppressed even below} $1/2$. This
relaxation--induced tendency to order of the electronic transfer
through the dot is unexpected, since we are dealing with an oscillator
with incoherent dynamics coupled to a SET in the sequential tunneling
regime. The onset of this behavior is discussed as a function of
relaxation, e--ph interactions, and external voltages.  We have found
that for {\em any} value of the e--ph coupling, a critical voltage
$V_{t}(\lambda)$ exists such that for $V>V_{t}(\lambda)$ a suppression
of the Fano factor below $1/2$ is possible.  At low voltages, these
results are qualitatively predicted by a four states toy model. In
this work we have focused on the case of a symmetric device on
resonance, which is the most favorable to analyze the suppression of
the Fano factor. However, qualitatively analogous results can be
obtained for asymmetric barriers and (or) $\bar{\varepsilon} \ne 0$.\\

Financial support by the EU via Contract No. MCRTN-CT2003-504574 and
by the Italian MIUR via PRIN05 is gratefully acknowledged.


\begin{thebibliography}{99}
\bibitem{weights} The general expression for the weight factors is:
 $ a_l=\exp[-\lambda^2 \coth(\beta \omega_0/2)] \exp[l \beta
\omega_0/2]I_l\left(\lambda^2/\sinh(\beta \omega_0/2) \right)$, where
$I_l$ is the modified Bessel function of first kind.
\end{thebibliography}
\end{document}